\documentclass[
showkeys,	
reprint,	
amsmath,	
amssymb,	
amsfonts,	
aps,		
prl%
groupedaddress,
superscriptaddress,
a4paper,%
nofootinbib,
eprint,		
final,		
preprintnumbers, 
]{revtex4-2}

\usepackage[tbtags]{mathtools}
\usepackage{amsmath,amsfonts,mathdots,amssymb,yfonts,calc}
\usepackage{graphicx,graphics,xcolor}
\usepackage{subfigure}
\usepackage{amsthm}

\usepackage{natbib}
\usepackage{booktabs}
\usepackage{siunitx}
\usepackage{float}
\usepackage{tabularray}
\usepackage{multirow}
\usepackage{physics}
\usepackage{dcolumn}
\usepackage[utf8]{inputenc}  
\usepackage[T1]{fontenc} 
\usepackage{natbib}
\usepackage{booktabs}
\usepackage{hyperref}
\hypersetup{
    colorlinks=true,
    linkcolor=blue,
    filecolor=magenta,      
    urlcolor=blue!80,
    citecolor=cyan,
    linktocpage=true,
    breaklinks=false,
}

\begin{document}
\title{Finite-scale geometric invariants for chaotic and weakly chaotic dynamics}
\author{Vinesh Vijayan}
\email[]{vinesh.physics@rathinam.in}
\affiliation{Department of Science and Humanities, Rathinam Technical Campus, Coimbatore, India -641021}

\date{\today}

\begin{abstract}
We introduce a finite-scale geometric observable that quantifies the growth rate of localized sets under time evolution in dissipative dynamical systems. Defined at finite time and resolution—without reference to symbolic dynamics or Markov partitions—this observable converges, in uniformly hyperbolic systems, to a resolution-dependent plateau whose logarithmic scaling coefficient equals the Kolmogorov-Sinai entropy. In merely hyperbolic systems, it decays to zero, reflecting the absence of entropy production, while remaining well-defined at finite scales. Numerical results for the Hénon map and Feigenbaum point illustrate these behaviors. Our findings yield a finite-scale geometric characterization of chaotic dynamics, consistent with classical entropy theory where applicable. We further demonstrate that the observable remains well defined in open intermittent systems, where trajectories escape and classical asymptotic invariants fail, revealing finite-scale signatures of transient weak chaos.
\end{abstract}

\maketitle
\pagestyle{plain}

\section{\label{s1}Introduction}
Chaotic dynamics is traditionally characterized by asymptotic quantities like Lyapunov exponents and Kolmogorov–Sinai (KS) entropy, which quantify exponential instability and information production. Complementary geometric views highlight the fractal structure of invariant sets, whose dimensions encode attractor complexity \citep{GrassbergerProcaccia1983PhysD, LedrappierYoung1985AnnalsI, LedrappierYoung1985AnnalsII}. Although Pesin's identity connects KS entropy to positive Lyapunov exponents in uniformly hyperbolic systems, both rely on infinite-time and infinitesimal-scale limits\citep{Pesin1977RMS,Ruelle1976,Ruelle1978,Ruelle1978thermodynform}.

In many physically relevant scenarios, chaos manifests at finite resolution and over finite times—especially in dissipative systems near crises or in experiments limited to coarse-grained geometry. Moreover, classical entropy concepts fail beyond uniform hyperbolicity, such as at critical attractors where Lyapunov exponents vanish and entropy production ceases\cite{Ott2002,GrebogiOttYorke1983PhysD}.

These limitations motivate observables that operate at finite scales and times, admit direct geometric interpretations, and recover known invariants in appropriate limits. Here, we introduce the finite-time geometric invariance rate—a measure of the logarithmic growth of localized sets under time evolution. Rather than extending entropy beyond its domain or using regime-specific quantities, our single observable reveals universality: in hyperbolic systems, its resolution dependence yields entropy as a scaling coefficient; in critical and transient regimes, it remains well-defined, capturing sub-exponential growth or escape rates instead. This finite-scale geometric perspective complements asymptotic theories. Finite-scale geometric signatures of chaos are increasingly relevant in experiments where resolution and observation time are intrinsically limited.\cite{Klein2020,Murali2021,Goldfriend2017,Boffetta2020}

Classical instability measures rely on asymptotic, closed-system dynamics, but many real systems are open, showing escape, decay, or transients without invariant measures—rendering Lyapunov exponents or entropies ill-defined\cite{OttYorke2008, Grebogi1983}. Transient dynamics can still exhibit geometric complexity and weak chaos, especially with marginal instabilities or intermittency\cite{Bonte2025}. This gap motivates a finite-scale geometric observable for instability at finite time and resolution, without assuming stationarity or exponential sensitivity.

\section{Finite-Scale Geometric Observable}
Consider the flow $\phi^t: M \to M$, a $C^{1+\alpha}$ diffeomorphism on a compact phase space $M$ preserving (or empirically sampled by) an invariant measure $\rho$. At finite numerical or experimental resolution $\epsilon$, the fundamental objects are finite neighborhoods, not infinitesimal displacements. We thus define the local unstable neighborhood
\begin{equation}
W^u_\epsilon(x) := W^u(x) \cap B_\epsilon(x),
\label{eq:E1}
\end{equation}
where $W^u(x)$ denotes the local unstable manifold of $x$ and $B_\epsilon(x)$  denotes the open metric ball of radius 
$\epsilon$ centered at x. A direct geometric measure of stretching in $W^u_\epsilon(x)$ is then $\mu(\phi^t(W^u_\epsilon(x)))$. The expansion ratio
\begin{equation}
g_\epsilon^t(x) := \frac{\mu(\phi^t(W^u_\epsilon(x)))}{\mu(W^u_\epsilon(x))}
\label{eq:E2}
\end{equation}
quantifies the relative growth of such sets---a finite-resolution analogue of exponential trajectory divergence.
Growth under composition is multiplicative:
\begin{equation}
\mu\bigl(\phi^{t+s}\bigl(W^u_\epsilon(x)\bigr)\bigr)
  \sim \mu\bigl(\phi^t\bigl(W^u_\epsilon(x)\bigr)\bigr)
     \cdot \mu\bigl(\phi^s\bigl(W^u_\epsilon(x)\bigr)\bigr),
\label{eq:E3}
\end{equation}
where on the right-hand side both factors use the same initial neighborhood \(W^u_\epsilon(x)\) for consistency.
Since growth under composition is multiplicative in time, the logarithm is the unique choice that makes growth additive. Taking the logarithm of Eq.~\eqref{eq:E2} yields an additive quantity, in terms of which the local growth rate for the neighborhood \(W^u_\epsilon(x)\) is
\begin{equation}
  \frac{1}{t} \log \frac{\mu(\phi^t(W^u_\epsilon(x)))}{\mu(W^u_\epsilon(x))} \,,
  \label{eq:E4}
\end{equation}
This local rate depends on the base point \(x\); averaging over the invariant measure \(\rho\) suppresses fluctuations at atypical points and defines the physically relevant geometric observable
\begin{equation}
  G(\epsilon,t) := \frac{1}{t} \int_M \log \left( \frac{\mu(\phi^t(W^u_\epsilon(x)))}{\mu(W^u_\epsilon(x))} \right) \, d\rho(x) \,.
  \label{eq:E5}
\end{equation}
By the Ledrappier--Young theory, for \(\mu\)-almost every \(x\),
\begin{equation}
  \mu(W^u_\epsilon(x)) = C(x) \, \epsilon^{d^u} \bigl[1 + o(1)\bigr] , \quad \epsilon \to 0 \,,
  \label{eq:E6}
\end{equation}
where \(d^u\) is the unstable dimension \cite{LedrappierYoung1985AnnalsI, LedrappierYoung1985AnnalsII,Ruelle1978}. Taking logarithms,
\begin{equation}
  \log \mu(W^u_\epsilon(x)) = d^u \log \epsilon + O(1) = -d^u \log(1/\epsilon) + O(1) \,.
  \label{eq:E7}
\end{equation}
The quantity \(G(\epsilon,t)\) in Eq.~(\ref{eq:E5}) is a rate per unit time, but its magnitude depends trivially on the resolution scale via Eq.~(\ref{eq:E7}). To compare growth rates across different resolutions, we must normalize by the logarithmic resolution units, \(\log(1/\epsilon)\). We therefore define a scale-regularized geometric growth rate
\begin{equation}
  I(\epsilon,t) := \frac{1}{t \, \log(1/\epsilon)} \int_M \log \left( \frac{\mu(\phi^t(W^u_\epsilon(x)))}{\mu(W^u_\epsilon(x))} \right) \, d\rho(x) \,.
  \label{eq:E8}
\end{equation}
With both geometric and dynamical origins, $I(\epsilon,t)$ represents the time-averaged logarithmic growth rate of localized sets---a finite-scale measure of expansion that is additive in time and independent of symbolic dynamics or partitions.
\subsection{ uniformly hyperbolic dynamics}
For uniformly hyperbolic dynamics, the evolution of the unstable neighborhood satisfies
\begin{equation}
  \mu(\phi^t(W^u_\epsilon(x))) \sim \epsilon^{d_u} \exp\left( d_u \int_0^t \lambda_u(\phi^s x)\,ds \right),
  \label{eq:E9}
\end{equation}
where \(\lambda_u\) is the local unstable Lyapunov exponent. Taking the logarithmic ratio from Eq.~(\ref{eq:E2}),
\begin{equation}
  \log \frac{\mu(\phi^t(W^u_\epsilon(x)))}{\mu(W^u_\epsilon(x))} =
  d_u \int_0^t \lambda_u(\phi^s x)\,ds + O(1).
  \label{eq:E10}
\end{equation}
Averaging over the invariant measure \(\rho\) (and using its stationarity) gives
\begin{equation}
  \int_M \log \left( \frac{\mu(\phi^t(W^u_\epsilon(x)))}{\mu(W^u_\epsilon(x))} \right) d\rho(x)
  = d_u \, t \int \lambda_u \,d\rho + O(1).
  \label{eq:E11}
\end{equation}
Substitution into the scale-regularized growth rate of Eq.~(\ref{eq:E8}) yields
\begin{equation}
  I(\epsilon,t) = \frac{d_u \int \lambda_u\,d\rho}{\log(1/\epsilon)}
  + O\left( \frac{1}{t \log(1/\epsilon)} \right).
  \label{eq:E12}
\end{equation}
By Pesin’s identity, \(d_u \int \lambda_u\,d\rho = h_{\rm KS}\), the Kolmogorov--Sinai entropy. Thus, in the long-time limit,
\begin{equation}
  \lim_{t \to \infty} I(\epsilon,t) = \frac{h_{\rm KS}}{\log(1/\epsilon)}.
  \label{eq:E13}
\end{equation}
Crucially, the entropy appears here as a scaling coefficient in the resolution dependence, not as a direct limit of the raw growth rate\citep{Pesin1977RMS, Ruelle1978}.

\subsection{Non-hyperbolic (critical) dynamics}

At a non-hyperbolic attractor, such as the Feigenbaum accumulation point of unimodal maps, the standard Lyapunov exponent vanishes:
\begin{equation}
  \lambda = \lim_{t \to \infty} \frac{1}{t} \log \| D\phi^t \| = 0,
  \label{eq:E14}
\end{equation}
so there is no exponential instability and the usual hyperbolic estimate fails. The diameter of a small unstable set is then controlled by the derivative,
\begin{equation}
  \mathrm{diam}(\phi^t(W^u_\epsilon)) \sim \| D\phi^t \| \, \mathrm{diam}(W^u_\epsilon).
  \label{eq:E15}
\end{equation}
Since initially \(\mathrm{diam}(W^u_\epsilon) \approx \epsilon\) and \(\| D\phi^t \| \sim t^\alpha\) for a wide class of critical maps, we obtain\cite{PomeauManneville1980}
\begin{equation}
  \mathrm{diam}(\phi^t(W^u_\epsilon)) \sim \epsilon \, t^\alpha.
  \label{eq:E16}
\end{equation}
Assuming self-similar scaling, the measure grows as
\begin{equation}
  \mu(\phi^t(W^u_\epsilon)) \sim \epsilon^{d_u} t^{\alpha d_u}.
  \label{eq:E17}
\end{equation}
It follows that
\begin{equation}
  \log \frac{\mu(\phi^t(W^u_\epsilon(x)))}{\mu(W^u_\epsilon(x))} = \alpha d_u \log t + O(1).
  \label{eq:E18}
\end{equation}
Substituting into Eq.~(\ref{eq:E8}) gives
\begin{equation}
  I(\epsilon, t) = \frac{\alpha d_u \log t}{t \log(1/\epsilon)} + O\left( \frac{1}{t \log(1/\epsilon)} \right),
  \label{eq:E19}
\end{equation}
so that
\begin{equation}
  \lim_{t \to \infty} I(\epsilon,t) = 0.
  \label{eq:E20}
\end{equation}
Thus, the invariant \(I(\epsilon,t)\) correctly signals the absence of entropy production at the critical point, in contrast to the hyperbolic regime where \(I(\epsilon,t)\) saturates to a finite resolution-dependent value proportional to \(h_{\rm KS}\).
\subsection{Crisis and transient chaos}
Following the destruction of a chaotic attractor at a boundary crisis, the dynamics is governed by a chaotic saddle and becomes effectively open. In this regime, the measure of trajectories remaining in the former attractor region decays exponentially with escape rate \(\kappa\),
\begin{equation}
  \mu(\phi^t(W^u_\epsilon)) \sim \epsilon^{d_u} e^{-\kappa t},
  \label{eq:E21}
\end{equation}
so that
\begin{equation}
  \log \frac{\mu(\phi^t(W^u_\epsilon(x)))}{\mu(W^u_\epsilon(x))} = -\kappa t + O(1).
  \label{eq:E22}
\end{equation}
Substituting into the scale-regularized geometric growth rate then yields
\begin{equation}
  \lim_{t \to \infty} I(\epsilon, t) = -\frac{\kappa}{\log(1/\epsilon)}.
  \label{eq:E23}
\end{equation}
\subsection{Finite-scale instability in open intermittent dynamics}
Intermittent dynamics generated by marginally unstable fixed points are known to produce subexponential instability and heavy-tailed escape statistics in
open systems~\cite{PomeauManneville1980,GeiselNierwetberg1982,OttGrebogiYorke1993}.
In this case, the finite-scale geometric invariant is defined in terms of the cumulative stretching along a trajectory,
\begin{equation}
\sum_{k=0}^{n-1} \log |Df(x_k)|,
\end{equation}
and does not depend on the existence of an invariant measure or asymptotic exponential growth.
In such systems, trajectories can remain close to marginal fixed points for extended periods, during which local expansion is weak. Thus, for trajectories surviving up to time $n$, the cumulative logarithmic stretching grows subexponentially and is expected to increase at most
logarithmically over finite, experimentally accessible times. Substituting this behavior into the general definition of the invariant yields
\begin{equation}
I(\epsilon,n)=
\frac{1}{n\log(1/\epsilon)}
\left\langle
\sum_{k=0}^{n-1} \log|Df(x_k)|
\right\rangle_{\mathrm{sur}},
\end{equation}
from which it immediately follows that the normalized quantity $n\log(1/\epsilon)\,I(\epsilon,n)$ remains bounded and independent of the resolution. In contrast to uniformly hyperbolic systems, no asymptotic plateau associated
with entropy production is expected; instead, subexponential instability generated by the marginal fixed point persists.
\section{Numerical Validations}
To demonstrate the scaling behavior of the finite-scale geometric invariant, we consider two paradigmatic low-dimensional maps that represent, respectively, hyperbolic and non-hyperbolic dynamics.
\subsection{Hyperbolic case- Henon map Example}
\begin{figure*}[hbt!]
    \centering
    \includegraphics[width=0.9\textwidth]{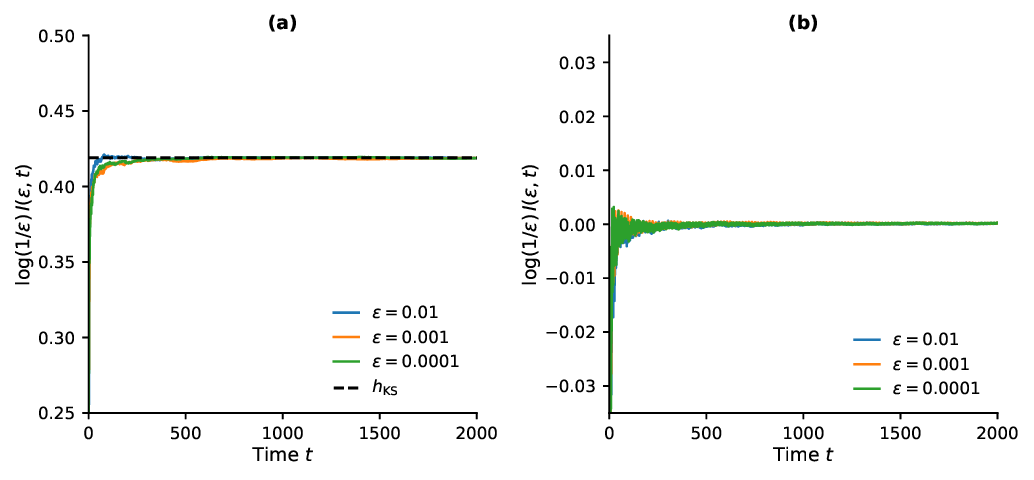}
    \caption{
    (a) Rescaled invariant $\log(1/\varepsilon)\,I(\varepsilon,t)$ for the Hénon map at $a=1.4$, $b=0.3$, showing convergence to the Kolmogorov--Sinai entropy $h_{\mathrm{KS}}$ (dashed line).
    (b) Same quantity for the logistic map at the Feigenbaum accumulation point $\mu=\mu_\infty$, where the invariant converges to zero for all resolutions.
    The same definition and normalization are used in both cases.
    }
    \label{fig:F1}
\end{figure*}
The uniformly hyperbolic case is illustrated by the H\'enon map\citep{Henon1976CMP}
\begin{equation}
  x_{n+1} = 1 - a x_n^2 + y_n, \quad y_{n+1} = b x_n,
  \label{eq:E24}
\end{equation}
with parameters \(a = 1.4\), \(b = 0.3\), where the attractor is chaotic and supports an SRB measure. For these parameters, the Kolmogorov--Sinai entropy is well established numerically as \(h_{\rm KS} \approx 0.419\). The invariant \(I(\epsilon,t)\) is evaluated by evolving tangent vectors under the Jacobian
\begin{equation}
  J(x,y) =
  \begin{pmatrix}
    -2a x & 1 \\
    b & 0
  \end{pmatrix},
  \label{eq:E25}
\end{equation}
and accumulating the geometric stretching along typical trajectories. Localized neighborhoods are aligned with the unstable direction, and their measure scales as \(\mu(W_\epsilon^u) \sim \epsilon^{d_u}\), with \(d_u = 1\) corresponding to the one--dimensional unstable manifold at each point of the attractor.

For the H\'enon map, trajectories are initialized randomly and iterated for a long transient to ensure convergence to the attractor. At each step, the Jacobian is applied to a unit tangent vector, and its norm gives the instantaneous stretching factor. The cumulative logarithmic growth is averaged over time and normalized by both the observation time $t$ and the logarithmic resolution scale $\log(1/\epsilon)$, exactly as prescribed by the invariant. Ensemble averaging is performed over many independent trajectories. No fitting parameters or post-processing corrections are introduced. In the chaotic regime of the H\'enon map, $I(\epsilon,t)$ converges at fixed resolution to a finite plateau whose value systematically depends on $\epsilon$. When multiplied by the resolution factor $\log(1/\epsilon)$, the curves for different $\epsilon$ collapse onto a common asymptote equal to the independently known Kolmogorov--Sinai entropy. This collapse, as depicted in Fig.~\ref{fig:F1}(a), confirms that the entropy appears as a scaling coefficient governing finite-scale geometric growth, rather than as a direct limit of the invariant itself. Residual fluctuations can be minimized by increasing the ensemble size.
\subsection{Non-hyperbolic case: Feigenbaum point}
The non-hyperbolic case is represented by the quadratic map at the Feigenbaum accumulation point\cite{Feigenbaum1978},
\begin{equation}
  x_{n+1} = 1 - r x_n^2,
  \label{eq:E26}
\end{equation}
with the critical parameter \(r_\infty = 1.401155189092\ldots\) corresponding to the onset of chaos via period doubling. At this critical parameter value, the Lyapunov exponent vanishes and instabilities grow subexponentially. The invariant is computed using the absolute value of the derivative,
\begin{equation}
  |f'(x)| = 2r|x|,
  \label{eq:E27}
\end{equation}
and the same regularization by \(\log(1/\epsilon)\) as in the hyperbolic case. The Feigenbaum attractor has a fractal structure, but for the present analysis only the logarithmic scaling and the local measure of unstable neighborhoods are needed. At the Feigenbaum point, the rescaled quantity converges to zero, as shown in Fig.~\ref{fig:F1}(b) for all resolutions studied. This behavior reflects the absence of exponential instability and is consistent with subexponential, algebraic growth of perturbations at criticality. 

The same numerical procedure and normalization are used in both cases, without any modification of the definition of the invariant. The contrasting asymptotic behavior arises solely from the difference in the underlying dynamical scaling, demonstrating that the invariant cleanly distinguishes hyperbolic and non-hyperbolic regimes through their finite-scale geometric structures.
\subsection{Open intermittent dynamics - Pomeau–Manneville map}
We consider the Pomeau-Manneville map\cite{PomeauManneville1980}, an open intermittent system on $x \in [0,1)$ defined by
\begin{equation}
x_{n+1} = x_n + a x_n^{z} \quad \text{for } z > 1,
\label{eq:28}
\end{equation}
with trajectories removed upon reaching $x \geq 1$. The marginally unstable fixed point at $x=0$ produces intermittent dynamics, featuring long laminar phases interrupted by bursts away from the neutral region. Consequently, the Lyapunov exponent vanishes, and exponential sensitivity to initial conditions is absent.
\begin{figure*}[t]
    \centering
    \includegraphics[width=0.6\textwidth]{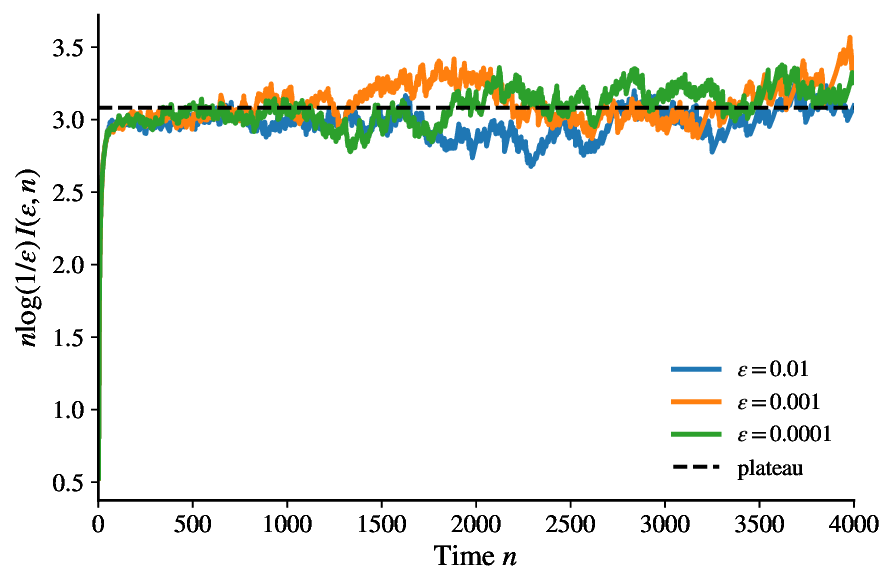}
    \caption{
    Finite-scale geometric invariant for an open intermittent map. Shown is the normalized quantity     $n\log(1/\varepsilon)\, I(\varepsilon,n)$ as a function of time $n$ for the open Pomeau--Manneville map $x_{n+1} = x_n + a x_n^{z}$ with $z=1.5$. Trajectories are removed upon reaching $x\ge1$, and averages are conditioned on survival up to time $n$. Curves corresponding to different resolutions $\varepsilon$ collapse onto a common bounded behavior, indicating subexponential instability characteristic of transient weak chaos. The dashed line marks the mean plateau value obtained from the smallest resolution.
    }
    \label{F2}
\end{figure*}
In such open systems, trajectories escape in finite time, with long-time statistics governed by the chaotic saddle. We thus use survival-conditioned averages, restricting to trajectories surviving up to the observation time. Substituting finite-time logarithmic stretching into the finite-scale invariant definition shows that the normalized quantity $n \log(1/\epsilon) I(\epsilon,n)$ remains bounded and independent of resolution $\epsilon$, as demonstrated in Fig.~\ref{F2}. The observed persistent fluctuations are expected, arising from long-tailed escape time distributions intrinsic to transient intermittent dynamics. Thus, the map provides a stringent test of the invariant in a regime where classical instability measures are zero or ill-defined.
\section{Conclusion}
The finite-scale geometric invariant introduced here quantifies the cumulative logarithmic stretching of localized sets under dynamical evolution. Without invoking symbolic dynamics or partition-based constructions, the invariant is defined at finite resolution and finite time, making it directly suitable for numerical simulations and experimental data. The normalization by the logarithmic resolution scale $\log(1/\varepsilon)$ follows naturally from the geometric scaling of phase-space structures and enables meaningful comparison of instability growth across different resolutions.

This single observable captures distinct dynamical regimes, as demonstrated by numerical results for the H\'enon map and the quadratic map at the Feigenbaum accumulation point. In uniformly hyperbolic systems, the invariant converges to
a resolution-dependent plateau whose logarithmic scaling coefficient coincides with the Kolmogorov--Sinai entropy $h_{\mathrm{KS}}$. For non-hyperbolic critical attractors, the invariant vanishes asymptotically, reflecting the absence of exponential instability and the emergence of subexponential, typically algebraic, divergence of nearby trajectories.

Beyond closed systems, we have shown that the same invariant remains well defined in open and transient dynamical regimes characterized by escape and
intermittency. In such systems, invariant measures may not exist and classical asymptotic indicators fail, yet survival-conditioned averages of the finite-scale invariant reveal bounded subexponential instability associated
with weak chaos. The observed behavior reflects the geometric structure of the underlying chaotic saddle rather than steady-state entropy production,demonstrating that meaningful instability information can be extracted even in
the presence of escape and strong temporal fluctuations.

Overall, the finite-scale geometric approach developed here complements classical asymptotic invariants such as Lyapunov exponents and Kolmogorov--Sinai entropy by providing a unified framework applicable to chaotic, critical, intermittent, and open dynamical systems. By characterizing instability directly at finite time and resolution, it bridges the gap between asymptotic theory and physically relevant finite-scale observations, extending the geometric understanding of chaos beyond the domain of uniform hyperbolicity.
\bibliography{reference}

\begin{thebibliography}{10}

\bibitem{GrassbergerProcaccia1983PhysD}
P.~Grassberger and I.~Procaccia, ``Measuring the strangeness of strange
  attractors,'' {\em Physica D: Nonlinear Phenomena}, vol.~9, no.~1-2,
  pp.~189--208, 1983.

\bibitem{LedrappierYoung1985AnnalsI}
F.~Ledrappier and L.-S. Young, ``The metric entropy of diffeomorphisms. {Part
  I}. characterization of measures satisfying {Pesin}'s entropy formula,'' {\em
  Annals of Mathematics}, vol.~122, no.~3, pp.~509--539, 1985.

\bibitem{LedrappierYoung1985AnnalsII}
F.~Ledrappier and L.-S. Young, ``The metric entropy of diffeomorphisms. {Part
  II}. relations between entropy, exponents and dimension,'' {\em Annals of
  Mathematics}, vol.~122, no.~3, pp.~540--574, 1985.

\bibitem{Pesin1977RMS}
Y.~B. Pesin, ``Characteristic {Lyapunov} exponents and smooth ergodic theory,''
  {\em Russian Mathematical Surveys}, vol.~32, no.~4, pp.~55--114, 1977.

\bibitem{Ruelle1976}
D.~Ruelle, ``A measure associated with axiom a attractors,'' {\em American
  Journal of Mathematics}, vol.~98, no.~3, pp.~619--654, 1976.

\bibitem{Ruelle1978}
D.~Ruelle, {\em Thermodynamic Formalism: The Mathematical Structures of
  Classical Equilibrium Statistical Mechanics}, vol.~5 of {\em Encyclopedia of
  Mathematics and its Applications}.
\newblock London: Addison-Wesley, 1978.

\bibitem{Ruelle1978thermodynform}
D.~Ruelle, ``Statistical mechanics of a one-dimensional lattice gas,'' {\em
  Communications in Mathematical Physics}, vol.~9, pp.~267--278, 1968.

\bibitem{Ott2002}
E.~Ott, {\em Chaos in Dynamical Systems}.
\newblock Cambridge: Cambridge University Press, 2nd~ed., 2002.

\bibitem{GrebogiOttYorke1983PhysD}
C.~Grebogi, E.~Ott, and J.~A. Yorke, ``Crises, sudden changes in chaotic
  attractors, and transient chaos,'' {\em Physica D: Nonlinear Phenomena},
  vol.~7, no.~1-3, pp.~181--200, 1983.

\bibitem{Klein2020}
M.~Klein, L.~Lepori, R.~Bachelard, and G.~Bachelard, ``Experimental measurement
  of entropy and lyapunov exponents in cold-atom systems,'' {\em Nature
  Physics}, vol.~16, pp.~1177--1182, 2020.

\bibitem{Murali2021}
S.~Murali, S.~Parthasarathy, S.~K. Senthilvelan, and M.~Lakshmanan,
  ``Finite-time chaos indicators in electronic circuits,'' {\em Physical Review
  Letters}, vol.~127, p.~154101, 2021.

\bibitem{Goldfriend2017}
T.~Goldfriend, A.~Kundu, J.~Avron, and Y.~Kafri, ``Experimental entropy
  production from coarse-grained dynamics,'' {\em Nature Physics}, vol.~13,
  pp.~1002--1006, 2017.

\bibitem{Boffetta2020}
G.~Boffetta, A.~Mazzino, and A.~Vulpiani, ``Finite-resolution chaos in
  turbulent flows,'' {\em Physical Review Letters}, vol.~125, p.~224101, 2020.

\bibitem{OttYorke2008}
W.~Ott and J.~A. Yorke, ``When lyapunov exponents fail to exist,'' {\em Phys.
  Rev. E}, vol.~78, no.~5, p.~056203, 2008.

\bibitem{Grebogi1983}
C.~Grebogi, E.~Ott, and J.~A. Yorke, ``Crises, sudden changes in chaotic
  attractors, and transient chaos,'' {\em Physica D: Nonlinear Phenomena},
  vol.~7, pp.~181--200, 1983.

\bibitem{Bonte2025}
M.~Bonte, V.~Lucarini, {\em et~al.}, ``Finite-size local dimension as a tool
  for extracting geometrical properties of attractors of dynamical systems,''
  {\em Nonlinear Processes in Geophysics}, vol.~32, pp.~139--162, 2025.

\bibitem{PomeauManneville1980}
Y.~Pomeau and P.~Manneville, ``Intermittent transition to turbulence in
  dissipative dynamical systems,'' {\em Communications in Mathematical
  Physics}, vol.~74, pp.~189--197, 1980.

\bibitem{GeiselNierwetberg1982}
T.~Geisel and J.~Nierwetberg, ``On the power spectrum of intermittent
  systems,'' {\em Physical Review Letters}, vol.~48, pp.~7--10, 1982.

\bibitem{OttGrebogiYorke1993}
E.~Ott, C.~Grebogi, and J.~A. Yorke, ``Chaotic scattering: An introduction,''
  {\em Physics Reports}, vol.~223, pp.~1--44, 1993.

\bibitem{Henon1976CMP}
M.~H{\'e}non, ``A two-dimensional mapping with a strange attractor,'' {\em
  Communications in Mathematical Physics}, vol.~50, no.~1, pp.~69--77, 1976.

\bibitem{Feigenbaum1978}
M.~J. Feigenbaum, ``Quantitative universality for a class of nonlinear
  transformations,'' {\em Journal of Statistical Physics}, vol.~19, no.~1,
  pp.~25--52, 1978.

\end{thebibliography}
\end{document}